\documentstyle[12pt,aaspp4]{article}
 
\lefthead{Hutchings et al.}
\righthead{Supersoft X-ray Binary CAL 87}
 
\begin{document}
\title{THE ECLIPSING SUPERSOFT X-RAY BINARY CAL 87}

\author{J.B. Hutchings\altaffilmark{1}, D. Crampton\altaffilmark{1}} 
\affil{Dominion Astrophysical Observatory\\Herzberg Institute of Astrophysics,
National Research Council of 
Canada\\ 5071 W. Saanich Rd., Victoria, B.C. V8X 4M6, Canada} 
\author{A.P. Cowley\altaffilmark{1}, \& P.C. Schmidtke\altaffilmark{1}} 
\affil{Physics \& Astronomy Dept., Arizona State University, Tempe, AZ, 
85287-1504} 

\authoremail{john.hutchings@hia.nrc.ca, david.crampton@hia.nrc.ca,
anne.cowley@asu.edu, paul.schmidtke@asu.edu} 

\altaffiltext{1} {Visiting Astronomers, Cerro Tololo Inter-American
Observatory, National Optical Astronomy Observatories, which is operated
by the Association of Universities for Research in Astronomy, Inc., under
contract with the National Science Foundation.}

\begin{abstract}

We present and discuss 25 spectra obtained in November 1996, covering all
phases of the CAL 87 binary system.  These spectra are superior both in
signal-to-noise and wavelength coverage to previously published data so
that additional spectral features can be measured.  Photometry obtained on
the same nights is used to confirm the ephemeris and to compare with light
curves from previous years.  Analysis of the color variation through the
orbital cycle has been carried out using archival MACHO
data.\footnote{This paper utilizes public domain data obtained by the
MACHO Project, jointly funded by the US Department of Energy through
Lawrence Livermore National Laboratory under contract W7405-Eng-48, the
National Science Foundation through the Center for Particle Astrophysics
of the University of California undercooperative agreement AST-8809616,
and the Mount Stromlo and Sidings Springs Observatory by the Bilateral
Science Technology and Regional Development.}  When a barely resolved red
field star is accounted for, there is no ($V-R$)-color variation, even
through eclipse.  There have been substantial changes in the depth of
minimum light since 1988; it has decreased more than 0.5 mag in the last
several years.  The spectral features and radial velocities are also found
to vary not only through the 0.44-day orbit but also over timescales of a
year or more.  Possible interpretations of these long-term changes are
discussed.  The 1996 spectra contain phase-modulated Balmer absorption
lines not previously seen, apparently arising in gas flowing from the
region of the compact star.  The changes in emission-line strengths with
orbital phase indicate there are azimuthal variations in the accretion
disk structures.  Radial velocities of several lines give different
amplitudes and phasing, making determination of the stellar masses
difficult.  All solutions for the stellar masses indicate that the
companion star is considerably less massive than the degenerate star.  The
Balmer absorption-line velocities correspond to masses of
$\sim$1.4M$_{\odot}$ for the degenerate star and $\sim$0.4M$_{\odot}$ for
the mass donor.  However, the strong He II emission lines indicate a much
more massive accreting star, with M$_X>$4M$_{\odot}$. 

\end{abstract}
 

\section{Introduction}

CAL 87 has long been known as one of only a small number of luminous X-ray
binaries in the Large Magellanic Cloud (Long, Helfand, \& Grabelsky 1981,
Pakull et al.\ 1988).  Its optical spectrum, with He II and H emission
lines on a very blue continuum, shows it to be similar to galactic
low-mass X-ray binaries.  However, its X-ray spectrum reveals CAL 87 to be
one of the rare, very luminous (L$_{bol}\geq10^{38}$ erg s$^{-1}$)
supersoft sources (SSS) which have little or no radiation above $\sim$0.5
keV (e.g. Tr\"umper et al.\ 1991, Greiner 1996).  The SSS are widely
thought to be binaries in which a white dwarf is undergoing rapid
accretion from a more massive companion, leading to steady nuclear burning
on the surface of the white dwarf (van den Heuvel et al.\ 1992). 
 
CAL 87 is unique among supersoft sources in having both optical and X-ray
eclipses (Callanan et al.\ 1989, Cowley et al.\ 1990: CSCH) which provide
extra information about the disk structure and in principle help to
constrain the stellar masses.  In the original spectroscopic data of CSCH,
He II 4686\AA\ emission was shown to move with K$=$40 km s$^{-1}$ and
proper phasing with respect to the eclipse so that it was interpreted as
due to orbital motion of the compact star.  CSCH concluded that the
compact star had a mass $\ge$6M$_{\odot}$, and hence these data implied
the presence of a black hole. 
 
However, optical spectra taken a few years later, combined with velocities
from lines in the far UV (Hutchings et al.\ 1995), showed a quite
different behavior.  The velocity amplitude was larger and the phasing was
very different, indicating that at times the velocities are not entirely
due to orbital motion and the line-formation regions change.  One problem
is that spectroscopic determination of the masses is not entirely
straightforward since the nearly edge-on view of the system
($i\sim78^{\circ}$) causes complications in the line profiles due to
motions in the accretion disk. 
 
In this paper we report on a series of spectroscopic data with improved
signal-to-noise (S/N) and new, concurrent photometry which we obtained at
CTIO in 1996 in order to conduct a more thorough investigation of CAL 87.

\section{Observations and Data}
\subsection{Spectroscopy}

CAL 87 was observed extensively during a five-night run in 1996 November
with the CTIO 4-m telescope using the KPGL1 grating and Loral 3K detector.
The wavelength range covered was $\sim$3700--6700\AA, with resolution
$\sim$3\AA.  Exposure times were 20 minutes, and the mean S/N was about 20
per resolution element.  The one-dimensional wavelength-calibrated spectra
were extracted and processed following standard IRAF procedures.
Twenty-five spectra were obtained, evenly distributed through the orbital
phases, with significantly better S/N and wavelength coverage than those
used in the CSCH paper.  Table 1 gives the journal of observations. 

Figure~\ref{med} shows the spectrum of CAL 87 in and out of eclipse, with
principal features identified.  The spectrum is similar to other supersoft
X-ray binaries, with the strongest features being emission lines of He II
and hydrogen.  However, the overall emission-line equivalent widths are
below typical values (see Cowley et al.\ (1998) for comparison with other
supersoft systems).  The He II Pickering-series emissions are sufficiently
strong that they must significantly contaminate the H Balmer emission
lines.  Thus, throughout this paper ``H$\alpha$", ``H$\beta$", etc.\ refer
to the blend of hydrogen and He II Pickering lines.  Also present are
emission lines of the high excitation ions O VI (3811, 3834, 5290\AA, also
seen in all other supersoft binaries), N V (4603, 4619\AA), and the C
III/N III blend ($\sim$4630-50\AA).  There is a broad emission at 4417\AA\
that we have been unable to identify.  It is similar in strength to He II
4541\AA.  The spectra also show Balmer absorption lines within the broad
emission profiles, not seen in previous spectra of CAL 87. 

The difference between the out-of-eclipse and in-eclipse spectra is also
shown at the bottom of Figure~\ref{med}.  One sees that the bluest part of
the continuum and some emission lines are eclipsed during each orbit.
However, in the photometry section below we will show that the apparent
reddening of the continuum in the $V$ and $R$ bands during eclipse,
reported by Alcock et al.\ (1997), is due to an unresolved red field star
which contributes about half of the light during eclipse. 

All of the individual spectra were examined as a function of time to see
if the observed changes are related to the well-known orbital period
(P=0.44 days) or to some other time scale.  It is clear that the spectral
changes are strongly phase-correlated, and we see no recognizable changes
on non-orbital timescales during the 5-night run.  On each individual
spectrum measurements were made of the radial velocity of He II 4686\AA\
and the line strengths of several lines.  In order to measure the weakest
features, the spectra were co-added in the observed wavelength frame in
eight phase bins, each with a spread of no more than 0.1P in phase.  The
phase 0 (eclipse) and phase 0.8--0.9 bins each contain four individual
spectra.  The phase 0.1 bin contains two spectra, and the remaining five
bins each contain three spectra.  Figure~\ref{binned} displays the
phase-binned spectra from $\sim$4100--4900\AA\ as well as the sum of the
spectra from phases 0.2--0.8.  Figure~\ref{bin2} shows more detail from
the phase-binned spectra in region from $\sim$4600--4900\AA.  Using these
spectra, radial velocities of He II 4686\AA\ and several other weaker lines
were measured independently by more than one author using different
techniques such as parabola-fitting, line centroiding, and
cross-correlating.  The agreement between sets of measures was well within
the estimated errors of $\pm5$ km s$^{-1}$ in velocity and 10\% in
equivalent width.  The mean values of all measures are listed in Table 1. 

However, there are complications in making the line measurements.  Some
emission lines are asymmetrical and are divided by absorption components
that move back and forth across them with phase (see Figure 3).  The
absorptions are most clearly seen at the H-Balmer lines but also may be
present in He II 4686\AA.  It is not clear whether the He II Pickering
lines are affected by absorption, but O VI, C III, and N V are probably
not.  In addition, the emission-line regions undergo different levels of
eclipse, and this affects their profiles and hence radial velocites during
the eclipse phases.  In our measures we attempted to separate the
absorption and emission components where possible.  Emission-line
velocities were based on the broad-line wings, and emission line strengths
do not include the absorption components.  However, where the absorption
and emission velocities are close, such separation is impossible
empirically.  Since the absorption strength changes with phase, modeling
by subtraction of a constant absorption profile is not feasible.  We
further discuss the consequences of these blends in a later section.
Figures~\ref{lines} and \ref{vels} show the phase variations of the
equivalent widths and radial velocities of selected lines, respectively. 

\subsection{Photometry}
  
To confirm the photometric ephemeris and look for possible changes in the
light curve, $V$-band CCD photometry of CAL 87 was obtained using the CTIO
0.9-m telescope in 1994 November and 1996 November.  The images were
calibrated with observations of Landolt's (1992) standard stars and
reduced with DAOPHOT (Stetson 1987), in a way similar to our earlier work.
Differential magnitudes were calculated for the $V$ filter relative to
local photometric standards within the CCD frames using a procedure which
minimizes errors by PSF fitting (Schmidtke 1988).  A line-of-sight field
star (separation $0.9^{\prime\prime}$, position angle $340^\circ$; Cowley
et al.\ 1991, Deutsch et al.\ 1996) was resolved on most images.  When
DAOPHOT could not separate the two stars, then the contribution of the
field star was subtracted from the blended light. 

The 1994 and 1996 photometry for CAL 87 is presented in Table 2.
Photometric phases have been calculated using the ephemeris of Schmidtke
et al.\ (1993), which accurately fits the new photometry and all other
data we have obtained over the past eleven years.  The ephemeris is: 

\vskip 5pt 
\centerline{T$_0 =$ HJD 2447506.8021($\pm0.0002$) $+$ 
N$\times$0.4426777($\pm0.0000016$) days} 
\vskip 5pt
\noindent
We note that the recent study of CAL 87 based on MACHO data obtained
between 1992 and 1996 finds virtually the same period and phasing which 
differs from ours by only $+$0.008P (Alcock et al.\ 1997). 

Figure~\ref{phot} shows our $V_{CTIO}$ photometry from all years of
observation.  A mean light curve, based on observations from 1985 November
through 1992 December (but with most of the data from 1988--1989), is
superposed.  Also shown is a plot of residuals of the individual data
points from this mean curve.  There is a marked change in the depth of the
primary minimum, particularly in the 1994 and 1996 observations which show
a decrease of $\sim$0.6 mag in $V$ (i.e. the eclipse became shallower).
The depth of secondary eclipse also varies, but to a lesser degree.  The
eclipse was deepest in 1988 and shallower at all other epochs for which we
have observations covering these orbital phases.  The out-of-eclipse light
curve does not show a corresponding brightening.  The same amount of extra
light which is present at primary mid-eclipse in 1996 would raise the
out-of-eclipse brightness by $\sim$0.15 mag, but our photometry shows no
change in the brightness outside of eclipse.  In the bottom panel of
Figure~\ref{phot} the dashed line shows the effect of adding this extra
source of light at \underbar{all} phases.  While this added light fits the
1994 and 1996 CTIO data quite well through eclipse, it is clear that this
amount of extra light is not present at other phases, as the majority of
points lie below the dashed curve. 

CAL 87 lies within one of the LMC fields monitored by the MACHO program.
To further investigate the change in the depth of eclipses, we obtained
public archival MACHO $V$ and $R$ data for CAL 87.  These data have been
discussed by Alcock et al.\ (1997) but with somewhat different objectives.
$V$ and $R$ magnitudes had already been calculated from the broad-band
red and blue colors in which the observations were made.  We do not know
exactly how the transformations were carried out, but we can compare the
CTIO and MACHO $V$ light curves since we have $V_{CTIO}$ data covering
similar dates.  We find that $V_{MACHO}$ is $\sim$0.15 mag brighter than
our $V_{CTIO}$ data \underbar{at} \underbar{all} \underbar{phases} in the
light curve, and we suggest that the problem may lie in the transformation
of the broad-band blue color to $V_{MACHO}$.  Unfortunately, we do not
have sufficient $R_{CTIO}$ data to compare the red light curves.  Thus,
for this part of the analysis we have modified our $V_{CTIO}$ magnitudes
by $-0.15$ mag in order to intercompare directly the two data sets and
study the changing eclipse depth. 

Adding all of the ($V-R$)$_{MACHO}$ data together, one derives a mean
color of ($V-R$)$_{MACHO}=-0.012\pm0.093$.  However, Deutsch et al.\ found
the out-of-eclipse color of CAL 87 to be somewhat redder, with
($V-R$)$_{HST}=+0.14$.  If, as suggested above, $V_{MACHO}$ is
systematically $\sim$0.15 mag too bright, then this would explain the
bluer mean color found in the MACHO data.  This also implies that
$R_{MACHO}$ is approximately similar to our $R_{CTIO}$ values (Cowley et
al.\ 1991). 

Figure~\ref{minJD} shows a plot of all MACHO data obtained within
$\pm0.03$P of central eclipse for both $V$ and $R$ light.  The modified
$V_{CTIO}$ minima ($V_{CTIO}-0.15$) observed between JD 2448900 and JD
2450300 are shown as crosses in Figure 7.  The plot shows that the depth
of minimum light has changed by almost a full magnitude during the $\sim$4
years of observation.  The depth of primary eclipse was greatest in 1988
and least in 1995.  Both $V$ and $R$ data, but especially $R$, suggest
that there may be a cycle of $\sim$600--700 days, although analysis shows
this variation is probably not strictly periodic, but rather only a
characteristic timescale for the changes.  Similar plots have been made of
the out-of-eclipse light, using data from restricted ranges of orbital
phase.  These show that the average light level remains unchanged through
the remainder of the orbit, consistent with the analysis of ``extra light"
discussed above and shown in Figure 6. 

One clue to understanding the change in the eclipse depth could lie in any
observed color changes.  Alcock et al.\ (1997) plot a ($V-R$)-color curve
which appears to show that CAL 87 reddens during eclipse, but since they
failed to correct their magnitudes for the effects of the red field star
(discussed above) which contributes nearly half the light of the system at
primary eclipse, their color curve is not correct.  Using the magnitudes of
Deutsch et al.\ (1996) which were derived from well-resolved HST images,
we have corrected the $V_{MACHO}$ and $R_{MACHO}$ magnitudes for the
presence of this nearby, unresolved star.  (Deutsch et al.\ found yet
another line-of-sight optical companion with a separation of
0.65$^{\prime\prime}$ in position angle 210$^{\circ}$.  However, it is
much fainter than the 0.9$^{\prime\prime}$ companion and has neglible
effect on the $V$ and $R$ light curves of CAL~87 at all phases.)  In our
analysis of 571 MACHO data points we have removed 24 individual points
with very large errors ($\ge0.4$ mag) and 7 which which lie $ge0.5$ mag
from the light curve, since these are obviously misidentifications or
blended images (CAL 87 lies in a very crowded field).  This resulted in
540 remaining ($V-R$) points well distributed in orbital phase.  We have
combined the data into 50 bins which are plotted versus phase in
Figure~\ref{color}.  The error bars shown are the standard error of the
mean.  It is clear that when the correction is made for the superimposed
field star there is \underbar{no} ($V-R$) color change with phase, even
during eclipse.  We have also examined the ($V-R$)-color curve for
individual years and for extreme eclipse-depth values.  Again, the
($V-R$)$_{MACHO}$ shows no variation with orbital phase.

If there is no change in the ($V-R$) color through the orbital cycle, one
might ask why the spectrum of the ``eclipsed flux" (as shown in Figure 1)
is bluer than the out-of-eclipse light.  In this paper we have only $V$
and $R$ photometry.  Since the hottest central regions of the disk are
eclipsed, it may well be that a ($U-B$) curve would show some variation
with phase, but we do not have enough $U$ or $B$ photometry to test this.
In principle we can use our spectra to measure color changes shortward of
V-band, provided we remove the field star contamination, which is small at
these wavelengths.  Doing so does suggest a small change in color in
eclipse, but accurate photometry is required to measure it. 

Considering that the depth of eclipse changes without a corresponding
change in the out-of-eclipse light and that there is no ($V-R$) color
change in the system through eclipse, the simplest explanation would be
that the shape of the eclipsing body is changing.  Since the companion
star is unlikely to change size, this idea then implies that the eclipse
is caused by an occulting region on the accretion disk rather than the
secondary star (or perhaps a combination of these two).  However, such a
scenario seems very unlikely, given the long-term stability of the eclipse
phasing.  A more plausible possibility is that part of the eclipsed light
comes from a hot spot on the outer part of the disk, as inferred in many
cataclysmic variables.  If the disk precesses, there may be periods when
this hot spot is hidden during eclipse, but later becomes visible through
eclipse when the angle between the disk and the observer has changed.  In
this picture the out-of-eclipse light would remain unchanged, as observed.

\section{Line Strengths and Radial Velocities}

In Figure~\ref{lines} we see that most lines show an increased equivalent
width through the eclipse, indicating that the lines are eclipsed less
than the continuum and thus must arise in an extended accretion disk.  The
exception is O VI 5290\AA, which becomes weaker near central eclipse,
indicating that it is eclipsed more than the continuum and hence is formed
in the innermost part of the disk. 

The emission-line strengths of He II and H show an asymmetry around
eclipse indicating extra flux is present in phases $\sim$0.7--0.9 (see
Figure~\ref{lines}).  The ratio of He II 4686\AA\ to H drops through these
phases, suggesting that the extra flux is greater in the Balmer lines than
in He II 4686\AA\ (or that more He II 4686\AA\ is hidden than H).  The O
VI line strength is unchanged over all phases other than during central
eclipse. 

The continuum is eclipsed by a factor $\sim$4.5, which is approximately
the equivalent-width increase of H$\alpha$ and H$\beta$.  Thus, the Balmer
emission region is essentially uneclipsed and must be very large with
respect to the secondary star.  He II 4686\AA\ is eclipsed to 50\% of its
out-of-eclipse level, and He II 5411\AA\ is eclipsed to $<$40\% of it
usual level.  The Pickering He II lines (as shown by 5411\AA) are eclipsed
over a shorter phase interval, and the eclipse profile is narrower than
for H and He II 4686\AA.  O VI is also eclipsed over a short interval to
$\sim$5\% of its out-of-eclipse level.  These changes imply that the
central hottest region, where O VI is formed, is eclipsed almost totally
while the Balmer emission mostly originates in a large region which is
only barely eclipsed. 

Best-fit sine curves for different sets of radial-velocity data are shown
in Table 3.  There appear to be several different phase-related velocity
variations.  Velocities that represent purely orbital motion of the
compact star should have maximum velocity at phase 0.75, assuming that
mid-eclipse corresponds to superior conjunction of the compact star.
However, the values in Table 3 clearly show that the situation is more
complicated, with variable line profiles and non-orbital motions
contributing to the observed velocities. 

In our original paper (CSCH) we used He II 4686\AA\ emission-line
velocities to derive the stellar masses, with the orbital inclination
being constrained by the light curve.  The velocities showed K$=$40 km
s$^{-1}$, with phasing which showed the emission came from the region
around the compact star.  These velocities implied a high-mass degenerate
star, much larger than a massive white dwarf.  The 1996 spectra reveal
that both the phasing and velocity amplitude of this line have changed, as
shown Table 3.  Depending on what part of the line is
measured (broad base or peak) the semi-amplitude lies in the range
$\sim26-33\pm5$ km s$^{-1}$, which implies even larger masses.  However,
the maximum velocity occurs at $\phi\sim0.6$ which is earlier than would
be expected if the velocities were only due to orbital motion.  If one
ignores the measurements near eclipse (phases 0.9--0.1), as there could be
some rotational distortion in the velocity curve due to different parts of
the disk being eclipsed at different phases, the maximum velocity moves to
phase $\sim$0.65, but the amplitude becomes even smaller, making derived
masses still higher.  However, the phasing indicates there is some
contribution to He II 4686\AA\ which is non-orbital and calls into
question whether we are able to determine the true orbital velocities and
hence derive stellar masses from velocities of this line.  We note that He
II 4686\AA\ shows a systemic velocity of $+$273 km s$^{-1}$, consistent
with CAL 87 being a member of the LMC. 

The He II 4686\AA\ line shows asymmetries that may be partially due to
superimposed weak absorption components, similar to those seen at the
hydrogen lines.  The H absorptions were much stronger in the 1996 spectra
than in previous years.  (He II 4686\AA\ differs from the He II
Pickering-series lines which do not show this asymmetry.)  A blended
absorption will have the greatest effect on the measured emission velocity
when it is strongest and off to one side.  This occurs at phases near
$\phi=0.6$ and $\phi=0.1$, so correction of the emission velocities for
the presence of absorption will have the effect of increasing the velocity
amplitude and causing the maximum velocity to occur later in phase. 

The sine-curve fit to the `H$\alpha$' emission velocities is shown in
Figure 5 and Table 3.  The amplitude is larger than for He II (K$=60$ km
s$^{-1}$) and the phasing later than expected from pure orbital motion.
From the variation in the strength of hydrogen emissions through eclipse
we have already determined that they are formed in a very large region,
and hence they are probably not good indicators of the compact star's
motion.  Notice that the systemic velocity for this line is $\sim$77 km
s$^{-1}$ more negative than that of He II, when the H$\alpha$ wavelength
is used to compute the velocity.  This is entirely consistent with the
line being a blend of H$\alpha$ and He II 6560\AA\ rather than pure
hydrogen. 

All of the O VI lines are too weak be measured on individual spectra, and
the 3800\AA\ lines are very noisy.  Measurements made on the phase-binned
spectra of O VI 5290\AA\ show a semi-amplitude K$=+35\pm26$ km s$^{-1}$,
but because the error is large this value should be taken with caution.  A
formal fit to velocities gives a maximum at phase $\sim0.83\pm$0.11 which,
given the large uncertainty, is consistent with the expected maximum at
$\phi=0.75$ for a line formed near the compact star.  Because the 5290\AA\
line appears to be blended with an unidentified feature on its
long-wavelength side, we have not been able to determine a reliable value
for its systemic velocity. 

The negatively displaced V$_0$ values of the Balmer absorption lines show
that they arise in gas flowing away from the system, with the velocity
component in our direction being $\sim$80 km s$^{-1}$ more negative than
the mean He II 4686\AA\ emission velocity.  Because we view the system
near to the disk plane, this outflow is in all azimuthal directions,
albeit with varying column density (see Figure~\ref{circle}).  Thus, we
appear to be viewing the system through a wind that has some azimuthal
density and velocity variation.  The absorbing medium evidently changes
over timescales of years, and this change may be connected with the
eclipse depth variations.  An azimuthally symmetrical wind velocity would
require the observed velocity change with phase to be orbital, but there
may be some wind velocity change with azimuth as well, although such a
wind would have to have spiral structure with respect to the disk. 
Because the absorption is weakest during eclipse, the absorbing gas must
be physically close to the disk.  The entire Balmer absorption phenomenon
appears to be different from what is seen in the supersoft X-ray binary
SMC 13 (Crampton et al.\ 1997), where the velocity amplitude is much
higher than can reasonably be due to orbital motion. 
 
The Balmer absorptions clearly have a larger velocity amplitude than the
emissions, as they move from one side to the other of the emission peaks.
The absorption velocities are similar for H$\beta$, H$\gamma$, and
H$\delta$, showing K=73 km s$^{-1}$ and phasing consistent with orbital
motion of the compact star.  However, Figure 5 shows the velocities are
very scattered, and with only six values it is very uncertain if they vary
smoothly through the orbital cycle. 

A very weak Ca II-K absorption line is seen in all phase-binned spectra.
It may also be formed in an outflow region, as V$_0$ is 40--70 km s$^{-1}$
lower than for He II emission.  Like H absorption, this feature was not
seen in spectra obtained in earlier years.  The mean Ca II-K equivalent
width ($\sim$1.1\AA) shows little change with phase, except for at
$\phi$=0 (EW=1.8\AA) and $\phi$=0.5 (EW=1.6\AA).  Thus, the Ca II
absorbing column does not change much around the orbit, but it is somewhat
higher near conjunctions.  The maximum velocity of Ca II-K absorption
occurs at an earlier phase ($\phi=$0.45) than for any other lines, and its
amplitude is the highest measured (K=97 km s$^{-1}$).  If the velocity
measured at the eclipse phase is ignored, then this result is even more
extreme (K$=$155 km s$^{-1}$) but with the same phasing.  The origin of
this line is not at all clear. 

Figure~\ref{circle} shows a sketch of the orbit plane and summarizes the
observed spectral changes.  The Balmer absorption appears to be strongest
in phases 0.05--0.23 and 0.6--0.7 and weakest at phase $\sim$0.4 and
during central eclipse.  The phasing of its velocity variations implies
association with the accretion disk, but the changing strength of the
absorption shows the column density varies with phase.  Since absorption
is not seen in the He II Pickering or other lines, their velocities should
be more reliable than the blended H and He II emission lines for
determining the motion of the compact star.  However, their weakness makes
them very difficult to measure reliably. 

\section{Discussion and Summary} 

We have discovered that the photometric eclipse depth has changed over
recent years, and this variation may even be cyclic.  The increase of
light observed at mid-eclipse may be due to a geometrical change in the
disk structure, without a significant change in total luminosity -- such
as a bright structure above or below the disk plane.  Alternatively, the
disk may grow in size in its plane, accompanied by extra absorption due to
the outflowing material, along lines-of-sight near the center.  The
emergence of the Balmer absorption in the 1996 spectra suggests the latter
scenario. 

In 1994 November we obtained three optical spectra of CAL 87 which were
reported by Hutchings et al.\ (1995) in their discussion of eight
ultraviolet spectra observed by HST in 1995 January.  We have remeasured
the optical spectra, although they are of considerably lower quality than
the new data and only show the strongest lines.  In 1994--5 He II 4686\AA\
and 1640\AA\ emission-line velocities show a very different phasing
(maximum velocity near $\phi\sim0.9$) and higher amplitude (K$\sim$70 km
s$^{-1}$) compared to both the CSCH data (K=40 km s$^{-1}$) and present
data (K$\sim$30 km s$^{-1}$).  The equivalent width of 4686\AA\ was lower
in the 1994 spectra than in 1996 by a factor of nearly two.  In two of the
three 1994 optical spectra weak absorption features are visible at
H$\beta$ and H$\gamma$, similar to those seen in 1996, but the low S/N
make the velocities unreliable.  Thus, it appears that there are long-term
spectral changes that may include significant non-orbital motions. 

The O VI lines might give a clean measure of orbital motion, as they arise
in the inner disk, but they are very weak, resulting in a large velocity
scatter.  The measured velocity amplitude (K=35 km s$^{-1}$) leads to the
same conclusion as CSCH, that the compact star must be massive, with
M$_X>$4M$_{\odot}$.  If instead the velocity amplitude of the compact star
is shown by the Balmer absorption lines (K=73 km s$^{-1}$), then the
resulting mass diagram is almost identical to that of the supersoft binary
SMC 13 (Crampton et al.\ 1997) and lower masses are determined.  We have
argued above that the measured He II velocity amplitude may be somewhat lower
than the actual velocity of the compact star, because the phasing of 
absorptions will decrease the apparent velocity extremes, while non-orbital
motions within the system add at other phases. We point out that the He II
velocities in CSCH, when the eclipse was deeper, gave K=40  km s$^{-1}$. 
The Balmer-line velocities are also likely to contain some non-orbital
motions.  Thus K=73 km s$^{-1}$ is a reasonable upper limit for the motion
of the degenerate star.  Figure~\ref{mass} shows the resulting masses
corresponding to these two extreme K values.  The masses are constrained
by the fairly well-known inclination ($i\sim70^{\circ}-80^{\circ}$) and by
the requirement that the mass-losing star fills its Roche lobe. 
No main sequence star fills the Roche lobe defined by these plots, for
orbital inclinations larger than $\sim$30$^o$. Thus, the mass-losing
star must have evolved off the main sequence to cause mass-exchange.
The lowest mass that can thus evolve within a Hubble time is $\sim$ 
0.4 M$_{\odot}$, and these are marked in the diagram at the appropriate
inclination value. These therefore correspond to the lowest X-ray star masses
under these assumptions, and just allow a white dwarf in the K=73 
km s$^{-1}$ case. It may be possible that the mass-losing star has very 
low mass by having lost most of it in some earlier event, and is still 
filling its Roche lobe now. However, in any of these cases the X-ray star 
is more massive and the resulting masses are not those expected by the
`standard' model (e.g. van den Heuvel et al.\ 1992) in which it is assumed 
that the compact star is a $\sim$1M$_{\odot}$ white dwarf and the donor 
star has a mass of $\sim$2.0 M$_{\odot}$. 

The spectrum in eclipse shows no signs of a late type spectrum (see
Figures~\ref{med},~\ref{binned},~\ref{bin2}), consistent with the
absence of an evolved more massive secondary.
Unfortunately, the complex changes in the disk spectrum make it difficult
to be more definitive than this at present, but it appears advisable to
revisit the evolutionary scenarios to accomodate a mass-losing star that
is less massive than the compact star, not only in CAL 87 but also for the
other supersoft X-ray binaries.  We discuss separately (Cowley et al.\
1998) the overall mass determinations for a number of supersoft binary
systems. 

Some of the supersoft X-ray binaries have been found to have highly
displaced lines indicating the presence of `jets'.  These are most
noticable in two low-inclination systems RX J0513$-$69 (Crampton et al.\
1996) and CAL 83 (Crampton et al.\ 1987) where the displacements are
several thousand kilometers per second.  In the intermediate inclination
system RX J0019$+$22 the jets show a velocity of $\sim\pm$800 km s$^{-1}$
from the central emission line.  The jet lines move with the same phase
and velocity amplitude as the central line, suggesting that the motion of
He II 4686\AA\ emission is indeed orbital.  Since CAL 87 is viewed nearly
edge-on, any line emission coming from such jets would have a very low
radial velocity and thus not be separated from the central emission
profile.  The asymmetry seen in the He II 4686\AA\ line of CAL 87 could
arise from a pair of shifted lines which have larger amplitude (or
possibly slightly different phase) than the central line.  This would
imply that the central line motion is not entirely orbital, but without
better resolution this suggestion cannot be verified. 

\acknowledgments

It is a pleasure to thank the CTIO staff for their assistance during our
observing runs.  We also thank Dr.\ Douglas Welch for considerable
assistance with the MACHO data.  APC gratefully acknowledges NSF support
for this work.

\clearpage

\centerline{Captions to Figures}

\figcaption[***.ps]{CAL 87 summed spectra in and out of eclipse and the
difference between these two spectra.  The eclipse-phase spectrum is from
the three observations taken between $\phi=$0.98 and 0.02.  Notice that
the Balmer absorption, seen superimposed on the emission in the upper
spectrum, is not present during eclipse.  Balmer emission is not
significantly eclipsed.  The eclipsed continuum appears to be fairly blue.
\label{med}} 

\figcaption[***.ps]{Phase-binned spectra showing the relative movement of
absorption and emission Balmer components.  Each spectrum is typically
summed from three individual spectra within 0.1P in phase (see text).
Vertical lines are at arbitrary wavelengths to enable visual alignment of
features with phase. \label{binned}} 

\figcaption[***.ps]{Phase-binned spectra showing asymmetry of He II
4686\AA\ and H$\beta$.  Spectra are spaced to allow easy viewing of the He
II line peaks. \label{bin2}} 

\figcaption[***.ps]{Variation of emission-line equivalent widths with
phase.  The peaks near $\phi_{phot}=$0 indicate that the emission lines
are not eclipsed as deeply as the continuum.  There is additional flux in
phases 0.7 to 0.9 in most lines.  O VI is more eclipsed than the continuum
and shows no extra flux in phases 0.7 to 0.9.  \label{lines}} 

\figcaption[***.ps]{Radial velocities of lines with phase.  Curves in all
panels show the best-fit sine curves to the points (see Table 3).  {\it
Upper panel:} Filled points are He II 4686\AA\ emission for the whole line
measured on individual spectra.  The open circles represent measurements
made in the phase-binned spectra.  {\it Middle panel:} Velocities of the
H$\alpha+$HeII emission line from the eight phase-binned spectra.  The
relatively low mean velocity of this feature is due to use of H$\alpha$
wavelength for the blend.  {\it Lower panel:} Velocity of hydrogen
absorption measured from H$\beta$, H$\gamma$, and H$\delta$ in the
phase-binned spectra. \label{vels}} 

\figcaption[***.ps]{The optical light curve of CAL 87. {\it Top panel:} A
composite curve based on CTIO $V$ data from 1985 to 1996, using the
ephemeris of Schmidtke et al.\ (1993).  The smooth curve represents the
mean behavior between 1985 and 1992, but the eclipse phases are dominated
by the 1988 points.  {\it Bottom panel:} Residuals from the mean curve,
showing the variable depth of primary eclipse which was noticably
shallower in 1994 and 1996.  The dashed line represents the expected
light-curve residuals if a constant extra-light contribution,
corresponding to 0.6 mag at minimum, is added at all phases.  The fact
that the majority of observations lie below this line shows the extra
light is not present outside of eclipse. \label{phot}} 

\figcaption[***.ps]{In-eclipse $V$ ({\it upper panel}) and $R$ ({\it lower
panel}) magnitudes using only orbital phases 0.97$\leq\phi\leq0.03$
(eclipse) versus Julian Date of observation.  Filled symbols are from
MACHO data; crosses are from our CTIO photometry.  All data have been
corrected for the 0.9$^{\prime\prime}$-separation red field star.  CTIO
$V$ magnitudes have also been adjusted by $-0.15$ mag, since the mean
out-of-eclipse $V_{MACHO}$ data are 0.15 mag brighter than $V_{CTIO}$ (see
text for discussion of this difference).  The symbol labeled `1988 range'
shows the range of magnitudes during phases 0.97--0.03 in the mean light
curve derived from 1988 data (i.e. the blur expected in observed minima).
Over the period of observation the depth of eclipse has varied by almost a
full magnitude.  No strict periodicity is present but the timescale of the
variation is about two years. \label{minJD}} 

\figcaption[***.ps]{($V-R$)-color of CAL 87 versus orbital phase, based on
MACHO data.  The colors have been corrected for the nearby red field star,
as described in the text.  There is no evidence for any color change
through eclipse or at any other phase in the orbit.  These data include
540 individual observations taken between 1992 and 1996. \label{color}} 

\figcaption[***.ps]{Sketch of the orbit plane illustrating the phases and
lines-of-sight for the main variations measured in the 1996 spectra.  The
`abs' refers to the Balmer absorption.  Roche lobe and disk sizes are
suggested, consistent with He II emission arising between the stars and O
VI near the disk center.  Radial velocity maxima are presumably affected
by non-orbital motions and extra emission on the disk-trailing side, as
expected for mass-transfer from the Roche lobe-filling companion star.
\label{circle}} 

\figcaption[***.ps]{Masses of component stars for the two extreme values
of velocity amplitudes discussed.  The eclipse indicates an inclination of
$70-80^{\circ}$.  The dots show the implied masses for the lowest possible
mass ($\sim$0.4M$_{\odot}$) of the donor star filling its Roche lobe.  The
mass of the accreting star rises rapidly for K values below 40 km
s$^{-1}$.  The absorption velocities are consistent with a massive white
dwarf, but the emission velocities indicate a more-massive compact object.
All cases require the donor star to be the less-massive component.
\label{mass}} 

\clearpage

\begin{deluxetable}{cccccccc}
\tablenum{1}
\tablecaption{CAL 87 Emission-line Measures}
\tablehead{\colhead{HJD} &\colhead{$\Phi$\tablenotemark{a}} 
&\colhead{RV$_{He II}$}
&\multicolumn{5}{c}{--------------------- Equivalent Width (\AA) 
---------------------}\\
\colhead{(2400000+)} &&\colhead{(km s$^{-1}$)} &\colhead{He II 4686} 
&\colhead{~~H$\alpha$~~} &\colhead{~~H$\beta$~~}
&\colhead{He II 5411} &\colhead{~~O VI 5290~~}}
\startdata
     50389.67&    0.33& 240&   12&  16   &2.8&    1.7&     1.8\nl
     50389.68&    0.37& 291&    9&  18   &1.1&    0.9&     1.9\nl
     50389.74&    0.49& 301&   13&  26   &1.8&    3.6&     1.5\nl
     50389.75&    0.53& 297&   12&  21   &1.2&    1.3&     2.5\nl
     50389.81&    0.65& 283&   12&  21   &0.6&    2.5&     1.0\nl
     50389.82&    0.69& 315&   11&  23   &2.1&    2.2&     0.8\nl
     50390.68&    0.61& 305&    9&  19   &2.3&    2.1&     1.6\nl
     50390.69&    0.64& 310&    9&  17   &1.8&    1.5&     1.5\nl
     50390.77&    0.81& 300&   14&  22   &4.7&    3.6&     1.2\nl
     50390.78&    0.85& 304&   16&  23   &4.5&    2.1&     0.9\nl
     50390.84&    0.98& 276&   24&  51   &10.5&   3.5&     0.4\nl
     50390.86&    0.02& 201&   21&  42   &9.3&    2.1&     0.1\nl
     50391.68&    0.87& 291&   16&  22   &5.2&    3.6&     1.6\nl
     50391.69&    0.90& 272&   15&  24   &6.0&    3.1&     2.5\nl
     50391.74&    0.01& 217&   19&  58   &9.2&    2.9&     0.3\nl
     50391.75&    0.04& 221&   15&  46   &5.6&    2.9&     0.3\nl
     50391.79&    0.13& 289&    9&  22   &2.1&    1.5&     1.3\nl
     50391.83&    0.21& 253&    9&  14   &1.1&    1.8&     1.4\nl
     50391.84&    0.25& 275&   12&  16   &3.0&    1.6&     1.4\nl
     50392.66&    0.09& 263&   12&  18   &2.8&    1.0&     1.5\nl
     50392.71&    0.21& 234&   11&  15   &2.6&    2.0&     1.0\nl
     50392.79&    0.38& 289&   11&  15   &2.0&    1.7&     1.2\nl
     50392.86&    0.54& 273&   11&  14   &2.3&    0.1&     1.1\nl
     50393.81&    0.69& 270&   11&  16   &2.6&    1.1&     1.5\nl
     50393.82&    0.72& 263&   12&  19   &4.0&    1.7&     1.8\nl
\enddata
\tablenotetext{a}{See text for ephemeris.}

\end{deluxetable}

\clearpage

\begin{deluxetable}{ccccccc}
\tablenum{2}
\tablecaption{CTIO Photometry of CAL 87 from 1994 and 1996}
\tablehead{\colhead{HJD} &\colhead{$V$}  &\colhead{$\sigma_V$}
 &\colhead{$\Phi_{phot}$}  \\
\colhead{(2400000+)}}
\startdata
49661.727 & 19.757 &  0.037 &  0.932 \nl
49661.743 & 20.095 &  0.030 &  0.967 \nl
49661.758 & 20.273 &  0.035 &  0.001 \nl
49661.772 & 19.952 &  0.023 &  0.034 \nl
49661.787 & 19.499 &  0.016 &  0.068 \nl
50389.799 & 18.894 &  0.025 &  0.631 \nl
50390.835 & 20.037 &  0.066 &  0.972 \nl
50391.708 & 19.777 &  0.048 &  0.945 \nl
50391.720 & 20.101 &  0.089 &  0.972 \nl
50391.732 & 20.197 &  0.076 &  0.998 \nl
50391.743 & 20.037 &  0.072 &  0.024 \nl
50391.755 & 19.766 &  0.044 &  0.050 \nl
50391.766 & 19.480 &  0.047 &  0.076 \nl
50392.719 & 18.893 &  0.031 &  0.228 \nl
50392.791 & 19.075 &  0.039 &  0.390 \nl
50393.714 & 19.061 &  0.038 &  0.477 \nl
50393.785 & 18.991 &  0.016 &  0.636 \nl
50393.854 & 19.310 &  0.022 &  0.792 \nl
50394.720 & 19.177 &  0.020 &  0.747 \nl
50394.816 & 19.962 &  0.053 &  0.966 \nl
\enddata

\end{deluxetable}

\clearpage

\begin{deluxetable}{lcccc}
\tablenum{3}
\tablecaption{Sinusoidal Fits to 1996 Spectroscopic Data}
\tablehead{\colhead{Data} &\colhead{V$_0$} &\colhead{K}
&\colhead{T$_0$} &\colhead{Mean error\tablenotemark{a}}\\
&\colhead{(km s$^{-1}$)} &\colhead{(km s$^{-1}$)} &\colhead{(phase)} 
&\colhead{(km s$^{-1}$)}}
\startdata

He II 4686\AA\ em -- wide(all) & 273$\pm$5  &  31$\pm$6 & 0.60$\pm$0.04 &  24\nl
 
He II 4686\AA\ em -- (all)     & 273$\pm$5  &  26$\pm$7 & 0.61$\pm$0.04 &  25\nl
 
He II 4686\AA\ em -- (8 bins)  & 273$\pm$6  &  33$\pm$8 & 0.62$\pm$0.05 &  18\nl
 
H$\alpha$+He II em -- (8 bins) & 196$\pm$8 & 60$\pm$13 & 0.82$\pm$0.03 &  24\nl
 
H abs -- (6 bins)      & 189$\pm$44 & 73$\pm$47 & 0.72$\pm$0.17 &  92\nl
 
O VI em -- (7 bins)           & - - & 35$\pm$26 & 0.83$\pm$0.11 &  46\nl
 
Ca II-K abs -- (8 bins)       & 232$\pm$34 & 97$\pm$51 & 0.45$\pm$0.08 &  96\nl
  
Ca II-K abs -- (7 bins, w/o $\phi$=0) & 205$\pm$25 & 155$\pm$40 & 
0.46$\pm$0.03 & 64\nl

He II em (CSCH, from 1989)  & 306$\pm$8 & 40$\pm$12 & 0.76$\pm$0.13 &  \nl

\enddata
\tablenotetext{a}{Average deviation of observations from fitted curve.} 

\end{deluxetable}
 
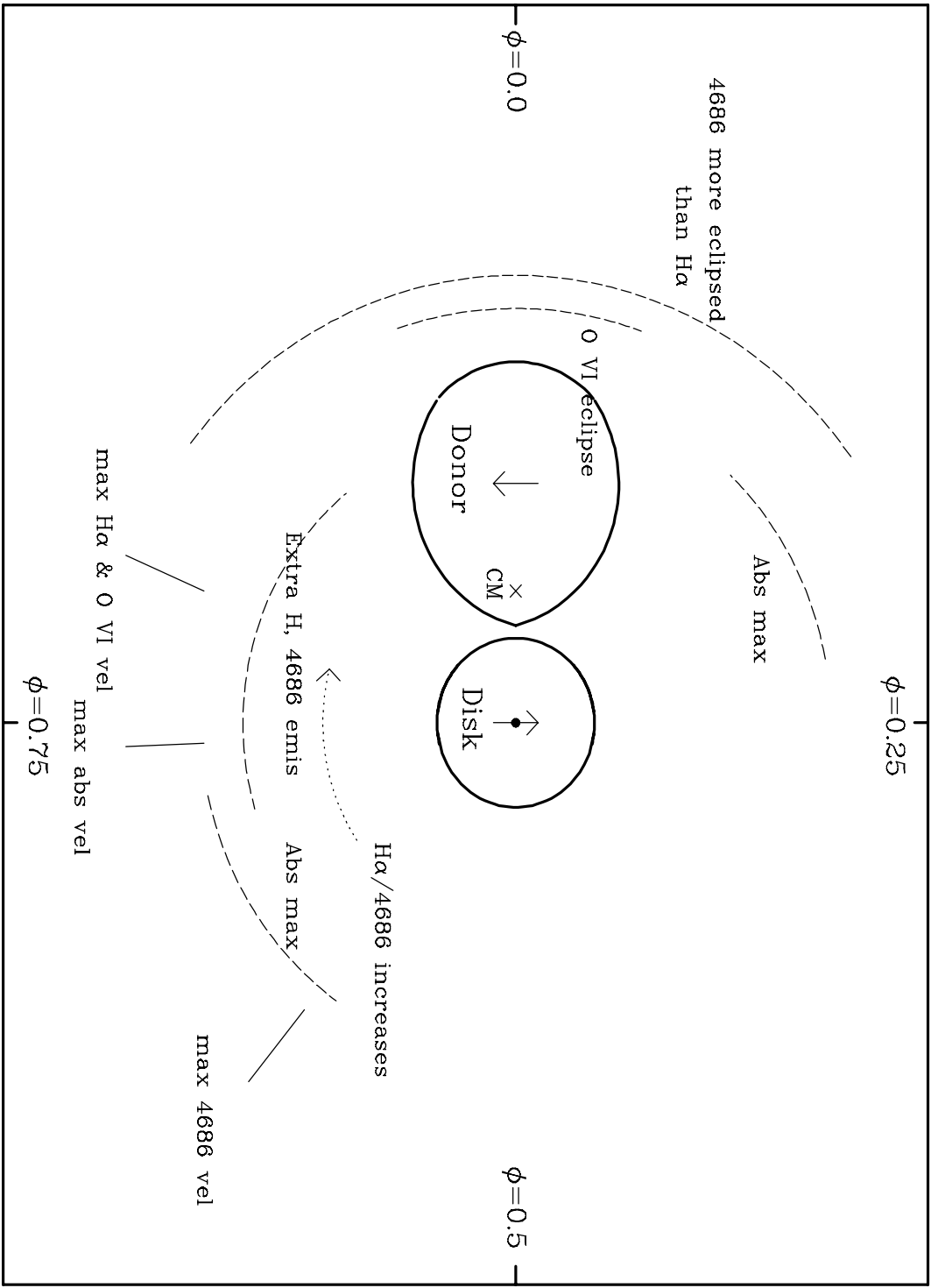
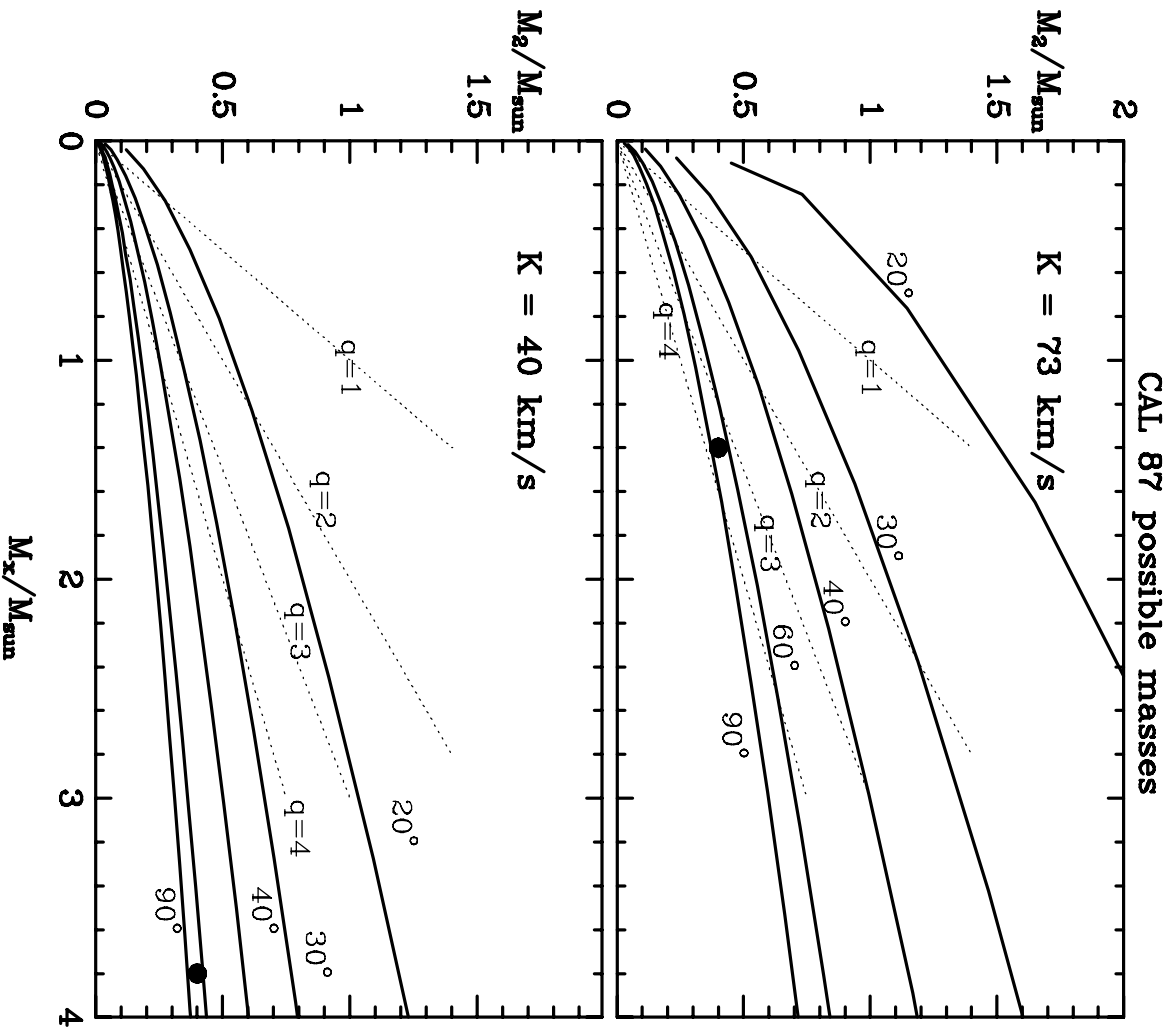
\end{document}